\documentclass[pre,twocolumn]{revtex4}
\usepackage{graphicx}
\usepackage{amsopn}
\usepackage{amsfonts}
\usepackage{latexsym}
\usepackage{bm,amssymb,amsmath}

\begin{document}

\title{Correlations of vorticity inside a coherent vortex}

\author{I.V.Kolokolov$^1 \, ^2$, V.V.Lebedev$^1\, ^2$, M.M.Tumakova$^1\, ^3$}

\affiliation{$^1$ Landau Institute for Theoretical Physics, RAS, \\
142432, Chernogolovka, Moscow region, Russia; \\
$^2$ National Research  University Higher School of Economics, \\
101000, Myasnitskaya ul. 20, Moscow, Russia.\\
$^3$ National Research  University Higher School of Economics, 190008,
 St.Petersburg, Russia}

\date{\today}

\begin{abstract}

We examine fluctuations of vorticity inside the coherent vortex, appearing as a consequence of the inverse energy cascade in two-dimensional turbulence. Temporal and spacial correlations can be characterized by the pair correlation function. The interaction between the fluctuations leads to non-zero value of  the third moment of vorticity. We examine the pair correlation function and the third moment for the model where the pumping is short correlated in time. We find explicit expressions for the Gaussian spacial correlation function of the pumping force. They confirm the general predictions obtained earlier.

\end{abstract}

\maketitle

\section{Introduction}
\label{sec:intro}

Two-dimensional turbulence is a subject of numerous investigations \cite{boffetta2012two}. The state can be observed in fluid films on scales larger than the film thickness. From the practical point of view, the most interesting such fluid ``film'' is atmosphere. Of course, atmosphere is very complex object and its detailed description is extremely difficult problem. However, some general features of two-dimensional turbulence could be useful for understanding atmospheric phenomena. Note in this respect the trend of the production of large-scale eddies from small-scale ones thanks to the non-linear hydrodynamic interaction in two-dimensional fluids \cite{67Kra,68Lei,69Bat}. In presence of the external forcing (pumping) the trend leads to formation of the inverse energy cascade at scales larger than the pumping length \cite{KraM}.

In a finite box, the transfer of the energy to large scales leads to formation of big vortices with the diameter of the order of the size of the box. Such vortices were observed both in laboratory experiments \cite{ShaF,FilLevchOrlov17} and in numerical simulations \cite{MCH04,chertkov2007dynamics}. At some conditions the big vortex has an infinite life time, that is it possesses long time correlations. We call such big vortices coherent. In the work \cite{laurie2014universal} the flat profile of the mean velocity of the coherent vortex was observed in numerical simulations and some arguments explaining the profile were presented. In the works \cite{KL15,kolokolov2016structure,kolokolov2016velocity,frishman17,kolokolov2017static} the quasilinear regime of the flow fluctuations inside the coherent vortex was ground, the flat velocity profile was derived and some velocity correlation functions were discussed. In the work \cite{KL20} the criterion of the formation of the coherent vortex was established, the criterion was confirmed in direct numerical simulations \cite{DFKL22}.

We consider the case where the two-dimensional turbulence is excited by a permanent small-scale pumping (external force). If the force is static or has statistical properties homogeneous in time then finally the statistically homogeneous in time turbulent state is achieved. Just in this situation the coherent vortices could appear. The laboratory experiments and the numerical simulations show that the coherent vortex is isotropic in average. The isotropy is explained by fast rotation suppressing the anisotropic perturbations on the background of the isotropic rotation. Let us stress that the mean flow of the coherent vortex is highly non-solid differential rotation. Such motion essentially deforms small-scale fluctuations, suppressing their nonlinear interaction. The property explains the applicability of the quasilinear approximation for the flow fluctuations inside the coherent vortex. The deformation makes correlation functions of the flow fluctuations highly anisotropic. The structure of the correlation functions is of crucial importance for such processes as transport of the passive scalar (the temperature field or the concentration of pollutants) inside the coherent vortex.

We are interested in the correlation functions of the fluctuating vorticity. In the quasilinear approximation the correlation functions are reduced to the pair correlation function. To find the correlation function, one has to examine the evolution of the small-scale flow fluctuations, produced by pumping, inside the coherent vortex. Some universal scaling properties of the correlation function were formulated in our work \cite{KLT23}. However, it is important to check the results by explicit calculations by using some concrete model. We chose the model where the pumping is assumed to be random processes short correlated in time. The spacial pair correlation function of the external force is assumed to be a Gaussian function of the distance between the points. The analytical calculations in the model enabled us to find explicit analytic expressions for the correlation function of vorticity, confirming the predictions of the work \cite{KLT23}.

Though the nonlinear interaction of the flow fluctuations inside the coherent vortex is weak, it plays an essential role in some processes. As an example, we note the energy transport inside the coherent vortex. In addition, the strength of the interaction grows as one goes away from the center of the vortex. Thus, evaluation of the strength is needed to establish the vortex boundary and the character of the fluctuations near the boundary. The strength of the interaction can be characterized by the third moment of the vorticity, that is zero in the quasilinear approximation. Some general properties of the moments of the vorticity were formulated in the work \cite{KL23}. However, again, it is important to check the results by explicit calculations by using some concrete model. We calculated the third moment in the framework of the above model and found the explicit analytic expression, confirming the predictions of the work \cite{KL23}.

The structure of our text is as follows.  In the section \ref{sec:general} we give some general relations concerning the fluctuations inside the coherent vortex. In the section \ref{sec:perturb} we evolve the perturbation theory needed for calculation of the correlation function of vorticity. In the section \ref{sec:pair} we examine the pair correlation function of vorticity. In the section \ref{sec:third} we calculate the third moment of vorticity. In the section \ref{sec:conclu} we sum up the results of our research.

\section{General relations}
\label{sec:general}

We consider two-dimensional turbulence excited by an external random force with a stationary statistics. We analyze the statistically stationary state realized after some transient process. In the state the characteristics of the coherent vortices are independent of time. Let us give two conditions needed for appearing the coherent vortices in the system. The first condition is
\begin{equation}
\epsilon \gg L^2 \alpha^3,
\label{criterion1}
\end{equation}
where $\epsilon$ is the energy production rate per unit mass (supplied by the external forcing), $L$ is the size of the box and $\alpha$ is the bottom friction coefficient. The condition (\ref{criterion1}) means that the energy carried by the inverse cascade accumulates at the scale $L$. The second condition is
\begin{equation}
\alpha \lesssim \nu k_f^2,
\label{criterion2}
\end{equation}
where $\nu$ is the kinematic viscosity coefficient and $k_f$ is characteristic wave vector of the external force. The condition (\ref{criterion2}) was derived in the work \cite{KL20} and confirmed by direct numerical simulations, see Ref. \cite{DFKL22}.

It is convenient to examine the coherent vortex in the reference frame attached to the vortex center. In the reference frame the mean (coherent) flow is differential rotation, which can be described by the polar (tangential) velocity $U$ depending on the distance $r$ from the observation point to the vortex center. Locally, the differential rotation is the shear flow with the shear rate
\begin{equation}
\Sigma=r \partial_r(U/r),
\label{shearrate}
\end{equation}
depending on $r$ as well. Note that for the solid rotation, where $U\propto r$, the shear rate (\ref{shearrate}) is zero. For the flat velocity profile, derived in the works \cite{KL15,kolokolov2016structure,kolokolov2017static}, $U$ is independent of $r$. Then the shear rate behaves as $\Sigma\propto r^{-1}$, that us $\Sigma$ diminishes as $r$ grows. Thus, the suppression of the interaction of the flow fluctuations, associated with the coherent vortex, slackens as $r$ increases.

The main object of our investigation are correlation functions of the vorticity $\varpi$, which fluctuates on the background of the average vorticity $\Omega$ in the coherent vortex. We assume that the fluctuations are produced by pumping, which has relatively small correlation length $k_f^{-1}$. At examining the fluctuations near a circle of radius $R$ much larger than the pumping correlation length, $k_f R\gg1$, we can limit ourselves to considering a narrow neighborhood of the circle. Then after passing into the reference system rotating with the angular velocity $\Omega(R)$ we obtain the following linear equation for $\varpi$
\begin{equation}
(\partial_t+\hat{\mathcal L})\varpi=\phi, \quad
\hat{\mathcal L}=
\Sigma x_2 \frac{\partial}{\partial x_1}
+\alpha -\nu \nabla^2,
\label{basic1}
\end{equation}
where $\Sigma$ is the shear rate (\ref{shearrate}), taken at $r=R$. The tangential variable $x_1=R\varphi$ ($\varphi$ is polar angle) and the radial variable $x_2=r-R$ are local Cartesian coordinates in the reference system. In the equation (\ref{basic1}) $\phi$ is curl of the external force $\bm f$ per unit mass, $\phi=\mathrm{curl}\, \bm f$.

The criterion justifying the quasilinear approximation is
\begin{equation}
\Sigma \gg (\epsilon k_f^2)^{1/3},
\label{criterion3}
\end{equation}
where the right part of the relation is Kolmogorov non-linear rate at the pumping scale. In turn, the quantity should be much larger than the viscous damping on the same scale:
\begin{equation}
(\epsilon k_f^2)^{1/3}\gg \nu k_f^2.
\label{criterion4}
\end{equation}
The criterion (\ref{criterion4}) is no other than the condition of large Reynolds number at the pumping scale. The condition is necessary for exciting turbulence. Combining the inequalities (\ref{criterion3},\ref{criterion4}), one finds that
\begin{equation}
\Sigma \gg \nu k_f^2,
\label{criterion5}
\end{equation}
inside the coherent vortex.

The equation (\ref{basic1}) demonstrates that the problem is reduced to the hydrodynamic motion on the background of a stationary shear flow. The presence of the shear flow breaks the spatial homogeneity. More precisely, the homogeneity is broken along the second axis. The fact makes complicated investigations of the solutions of the equation (\ref{basic1}). The equation (\ref{basic1}) without pumping $\phi$ was analyzed in the work \cite{souzy}. We are interested in the correlation functions of $\varpi$. To find the correlation functions one should solve the equation (\ref{basic1}) for arbitrary $\phi$ and then average the corresponding product over the statistics of $\phi$.

In the work we use the model of pumping short correlated in time. Then statistical properties of pumping are determined by its pair correlation function
\begin{equation}
\langle\phi(t_1,\bm x)\phi(t_2,\bm y)\rangle
=-2\epsilon\delta(t_1-t_2) \nabla^2 \Xi(\bm x-\bm y).
\label{basic2}
\end{equation}
The angular brackets in Eq. (\ref{basic2}) and below designate mean values obtained by averaging over time in experiment. In the theoretical framework the angular brackets mean averaging over the system statistics. Laplacian $\nabla^2$ in Eq. (\ref{basic2}) is explained by the relation $\varpi=\mathrm{curl}\, \bm v$.

Further we exploit the particular shape of the pair correlation function of pumping
\begin{equation}
\Xi(\bm x)=\exp(-k_f^2 |\bm x|^2/2) .
\label{pumpg1}
\end{equation}
Then $\epsilon$ in Eq. (\ref{basic2}) is just the pumping rate (energy production rate per unit mass). The pumping correlation function (\ref{pumpg1}) enables us to find explicitly the pair correlation function $\langle \varpi \varpi \rangle$ and the third moment $\langle \varpi^3 \rangle$. However, their qualitative properties are universal and are independent of the particular shape of the pair correlation function of pumping.

In the work \cite{KLT23} the time $\tau_\star$ is introduced, determining the characteristic time, at which the second moment of $\varpi$ is formed. The time is defined as
\begin{equation}
\tau_\star =\left(\Sigma^2 \nu k_f^2\right)^{-1/3}, \quad
\Sigma \gg \tau_\star^{-1}\gg \nu k_f^2.
\label{taustar}
\end{equation}
The inequalities in Eq. (\ref{taustar}) are explained by the condition (\ref{criterion5}). Note also that $\alpha \tau_\star \ll 1$. The inequality is explained by the same condition (\ref{criterion5}) and the condition (\ref{criterion2}) needed for realizing the coherent vortex.

If one is interested in effects related to the interaction of the fluctuations then one should introduce the nonlinear term into the linear equation (\ref{basic1}). It is determined by the nonlinear term in Navier-Stokes equation, responsible for the interaction of the fluctuations. Adding the nonlinear term to the linear equation (\ref{basic1}), one finds
\begin{equation}
\partial_t\varpi + \hat{\mathcal L}\varpi
+\nabla\left(\bm v \varpi-\langle \bm v \varpi \rangle\right)=\phi,
\label{vactt}
\end{equation}
where $\bm v$ is the velocity of the fluctuating flow.

An introduction of the average $\langle \bm v \varpi \rangle$ into Eq. (\ref{vactt}) is related to the fact that we assume $\langle \varpi\rangle =0$,  $\langle \phi \rangle=0$. The first condition is related to the definition of the fluctuating flow. The second condition means that the external forcing does not produce vorticity in average. The term $\langle \bm v \varpi \rangle$ depends on coordinates inside the coherent vortex and, consequently, its divergence is non-zero. It is the reason to include the term into Eq. (\ref{vactt}).

Due to the incompressibility condition $\nabla \bm v=0$ it is possible to introduce the stream function $\psi$, related to the velocity components and to the vorticity as
\begin{equation}
v_1=\frac{\partial \psi}{\partial x_2}, \quad
v_2=-\frac{\partial \psi}{\partial x_1}, \quad
\varpi=-\nabla^2 \psi.
\label{stre1}
\end{equation}
To restore the velocity $\bm v$ from the vorticity field $\varpi$, one has to solve the Laplace equation $\nabla^2 \psi=-\varpi$ and then to calculate the velocity components in accordance with Eq. (\ref{stre1}). Let us stress that the stream function $\psi$, introduced by Eq. (\ref{stre1}), is related to the fluctuating part of the flow.

In the quasilinear approximation the third order moment $\langle \varpi^3 \rangle$ is zero. To find a non-zero contribution to the moment $\langle \varpi^3 \rangle$, one has to take into account the interaction between the flow fluctuations related to the non-linear term in Navier-Stokes equation. In our setup one has to use the non-linear equation (\ref{vactt}) where the interaction is described by the second-order term $\bm v \nabla \varpi$.

All our calculations are performed at the condition $\Sigma>0$. However, the results can be easily extended to negative $\Sigma$. Say, the simple symmetry reasoning show that the third moment $\langle \varpi^3 \rangle$ changes its sign at $\Sigma \to-\Sigma$. As to the pair correlation function $\langle \varpi(t,\bm x) \varpi (0,\bm y)\rangle$, the transformation $\Sigma\to - \Sigma$ means that one should change the sign of the second component $x_2\to-x_2$, $y_2\to -y_2$ without touching $x_1,y_1$.

\section{Perturbation theory}
\label{sec:perturb}

The interaction of the flow fluctuations can be consistently examined in terms of Wyld diagrammatic technique \cite{Wyld61}, based on the dynamic equation (\ref{vactt}). The diagram technique can be derived from the representation of the correlation functions of the vorticity $\varpi$ as functional integrals over $\varpi$ and an auxiliary field $\mu$ with the weight $\exp(-{\mathcal I})$ \cite{MSR73}, where
\begin{eqnarray}
{\mathcal I}={\mathcal I}_2+{\mathcal I}_{int},
\label{gener1} \\
{\mathcal I}_2= \int dt\, d^2x\, \mu (\partial_t+\hat{\mathcal L})\varpi
\nonumber \\
+\epsilon \int dt\, d^2x\, d^2 r\, \nabla^2\Xi(\bm x-\bm r) \mu(t,\bm x) \mu(t,\bm r),
\label{gener2} \\
{\mathcal I}_{int}=
\int dt\, d^2x\, \mu \bm v \nabla \varpi .
\label{generint}
\end{eqnarray}
Here the velocity $\bm v$ is implied to be expressed via the vorticity $\varpi$. Details of the technic can be found in the review
\cite{HRS}.

The pair correlation function of the vorticity is written as the following functional integral
\begin{equation}
\langle \varpi(t, \bm x) \varpi (0, \bm y)\rangle
= \int D\varpi\, D\mu\, e^{-{\mathcal I}}
 \varpi(t, \bm x) \varpi (0, \bm y),
 \label{paircofu}
\end{equation}
We introduce also the following pair average
\begin{equation}
\langle \varpi(t, \bm x) \mu (0, \bm y)\rangle
= \int D\varpi\, D\mu\, e^{-{\mathcal I}}
 \varpi(t, \bm x) \mu (0, \bm y).
 \label{greenofu}
\end{equation}
This correlator (\ref{greenofu}) is the Green function, since it determines the response of the system to an additional external force. Note that the average $\langle \mu \mu \rangle$ is zero.

One can evolve the perturbation theory for any correlation function expanding the weight $\exp(-{\mathcal I})$ in the third order term ${\mathcal I}_{int}$ (\ref{generint}) and calculating the resulting Gaussian functional integrals. The integrals are expressed in terms of the ``bare'' correlation functions determined by the quadratic term (\ref{gener2}):
\begin{eqnarray}
\langle \varpi(t,\bm x) \mu(0,\bm y)\rangle_0
=\int {\mathcal D}\varpi {\mathcal D}\mu\, e^{-{\mathcal I}_2}
\varpi(t,\bm x) \mu(0,\bm y),
\label{gener3} \\
\langle \varpi(t,\bm x) \varpi(0,\bm y)\rangle_0
=\int {\mathcal D}\varpi {\mathcal D}\mu\, e^{-{\mathcal I}_2}
\varpi(t,\bm x) \varpi(0,\bm y).
\label{genar3}
\end{eqnarray}
The average of the type $\langle \varpi \dots \mu \dots\rangle_0$ is determined by Wick theorem \cite{Wick50})and it is equal to the sum of products of the pair averages  (\ref{gener3},\ref{genar3}) organized by all possible pairings. Each term of the perturbation series corresponds to a Feynman diagram.

The expressions (\ref{gener2},\ref{gener3}) lead to the following equation for the bare Green function
\begin{equation}
(\partial_t+\hat{\mathcal L})
\langle \varpi(t,\bm x) \mu(0,\bm y)\rangle_0
=\delta(t)\delta(\bm x-\bm y).
\label{greenf}
\end{equation}
Remind that any Green function is zero at negative times due to causality. Therefore $\langle \varpi(t,\bm x) \mu(0,\bm y)\rangle$ is zero at $t<0$. The expression for the ``bare'' pair correlation function can be derived using Eq. (\ref{genar3}):
\begin{eqnarray}
\langle \varpi(t,\bm x) \varpi(0,\bm y)\rangle_0
=-2\epsilon\int d\tau \int d^2r d^2 z
\nabla^2\Xi(\bm r-\bm z)
\nonumber \\
\langle \varpi(t,\bm x) \mu(\tau,\bm r)\rangle_0
\langle \varpi(0,\bm y) \mu(\tau,\bm z)\rangle_0 . \quad
\label{genar4}
\end{eqnarray}
The correlation function is proportional to the energy production rate $\epsilon$, as it should be.

Due to the presence of the shear flow the space homogeneity is broken. Therefore the pair correlation functions depend generally on both coordinates. In the case Fourier transform should be performed over both coordinates. We define Fourier transform for the Green function and the pair correlation function as follows
\begin{eqnarray}
\langle \varpi(t,\bm x) \mu(0,\bm y)\rangle_0
=\int \frac{d^2k\, d^2q}{(2\pi)^4}
e^{i\bm k \bm x-i\bm q \bm y}
{\mathcal G}(t,\bm k,\bm q),
\label{fouri1} \\
\langle \varpi(t,\bm x) \varpi(0,\bm y)\rangle_0
=\int \frac{d^2k\, d^2q}{(2\pi)^4}
e^{i\bm k \bm x+i\bm q \bm y}
{\mathcal F}(t,\bm k,\bm q).
\label{fouri2}
\end{eqnarray}
The definition leads to the rules convenient for us.

For the Fourier transform ${\mathcal G}(t,\bm k,\bm q)$ (\ref{fouri1}) we derive from Eq. (\ref{greenf}) the following differential equation
\begin{eqnarray}
\left(\partial_t - \Sigma k_1\frac{\partial}{\partial k_2}
+\alpha+\nu k_1^2+\nu k_2^2\right){\mathcal G}(t,\bm k,\bm q)
\nonumber \\
(2\pi)^2\delta(t)\delta(\bm k-\bm q).
\label{gref1}
\end{eqnarray}
Since the equation (\ref{gref1}) is of the first order, it can be easily solved by the method of characteristics to obtain
\begin{eqnarray}
{\mathcal G}(t,\bm k,\bm q)
=(2\pi)^2 \theta(t) \delta(k_1-q_1)
\delta\left(k_2-q_2+\Sigma k_1t\right)
\nonumber \\
\exp\left(-\alpha t-\nu q_1^2 t -\nu q_2^2 t
+\nu\Sigma q_2 q_1 t^2-\frac{1}{3}\nu \Sigma^2 q_1^2 t^3
\right),
\label{gref2}
\end{eqnarray}
where $\theta(t)$ is the Heaviside step function.

\begin{figure}
\includegraphics[width=\columnwidth]{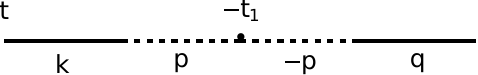}
\caption{Feynman diagram representing the pair correlation function (\ref{fouri4}).}
\label{fig:paircf}
\end{figure}

We find from Eq. (\ref{pumpg1})
\begin{eqnarray}
\int d^2 x\, \exp(-i\bm k \bm x) \Xi(\bm x)
= \frac{2\pi}{k_f^2}
\exp\left(-\frac{\bm k^2}{2k_f^2}\right).
\label{fouri3}
\end{eqnarray}
Therefore we obtain for the pair correlation function (\ref{genar4})
\begin{eqnarray}
{\mathcal F}(t, \bm k, \bm q)=
2\epsilon\int dt_1 \int \frac{d^2 p}{2\pi k_f^2}
p^2 \exp\left(-\frac{p^2}{2k_f^2}\right)
\nonumber \\
{\mathcal G}(t+t_1,\bm k,\bm p)
{\mathcal G}(t_1,\bm q,-\bm p).
\label{fouri4}
\end{eqnarray}
The expression (\ref{fouri4}) can be represented as the Feynman diagram depicted in Fig. \ref{fig:paircf}. Here the combined solid-dashed line represents the Green function (\ref{fouri1}), the bullet represents the pumping, and the integration over the wave vector $\bm p$ is implied.

To find a non-zero contribution to the moment $\langle \varpi^3 \rangle$, one has to take into account the interaction between the flow fluctuations described by the third order term ${\mathcal I}_{int}$ (\ref{generint}). The leading contribution to the third order moment is of the first order in the term:
\begin{equation}
\langle \varpi^3 \rangle
= -\int {\mathcal D}\varpi\, {\mathcal D}\mu\, \exp(-{\mathcal I}_2)
\varpi^3 {\mathcal I}_{int} .
\nonumber
\end{equation}
Using Wick theorem, one obtains
\begin{eqnarray}
\langle \varpi^3 \rangle
=-6 \int dt\, d^2x\,
\langle \varpi(0,\bm 0) \mu(t,\bm x) \rangle_0
\nonumber \\
\frac{\partial}{\partial \bm x}\langle \varpi(0,\bm 0) \varpi(t,\bm x) \rangle_0
\langle \varpi(0,\bm 0) \bm v(t,\bm x) \rangle_0.
\label{gener4}
\end{eqnarray}
In terms of the stream function $\psi$, see Eq. (\ref{stre1}), one can rewrite Eq. (\ref{gener4}) as
\begin{eqnarray}
\langle \varpi^3 \rangle
=6 \int_0^{\infty} dt\, \int dx_1 dx_2
\langle \varpi(t,\bm 0) \mu(0,\bm x) \rangle_0
\nonumber \\
\left[- \frac{\partial}{\partial x_1}\langle \varpi(t,\bm 0) \varpi(0,\bm x) \rangle_0
 \frac{\partial}{\partial x_2}\langle \varpi(t,\bm 0) \psi (0,\bm x) \rangle_0 \right.
\nonumber \\
\left. +\frac{\partial}{\partial x_2}\langle \varpi(t,\bm 0) \varpi(0,\bm x) \rangle_0
 \frac{\partial}{\partial x_1}\langle \varpi(t,\bm 0) \psi (0,\bm x) \rangle_0 \right].
\label{generr}
\end{eqnarray}
Here we have taken into account that the correlation functions depend on the time differences and that the Green function is zero at negative times $t$ due to causality.

\section{Pair correlation function}
\label{sec:pair}

The pair correlation function (\ref{fouri2}) can be examined using the obtained expressions. Substituting the expression (\ref{fouri4}) into Eq. (\ref{fouri2}) we find
\begin{eqnarray}
\langle \varpi(t,\bm x) \varpi(0,\bm y)\rangle_0
=2\epsilon\int_0^\infty dt_1 \int \frac{d^2 p\, p^2}{2\pi k_f^2}
\nonumber \\
\exp\left(-\frac{p^2}{2k_f^2}+W\right),
\nonumber \\
W=-\alpha (t_1+t_3)-\nu p_1^2 (t_1+t_3) -\nu p_2^2 (t_1+t_3)
\nonumber \\
+\nu\Sigma p_2 p_1 (t_1^2+t_3^2)-\frac{1}{3}\nu \Sigma^2 p_1^2 (t_1^3+t_3^3) +i p_1 x_1
\nonumber \\
+i(p_2-\Sigma p_1t_3)x_2
-i p_1 y_1-i (p_2-\Sigma p_1t_1) y_2 .
\nonumber
\end{eqnarray}
where $t>0$, $t_3=t+t_1$. The expression can be rewritten as
\begin{eqnarray}
\langle \varpi (t, \bm x) \varpi(0,\bm y)\rangle_0
=2\epsilon  k_f\frac{\partial}{\partial k_f}\int \frac{dp_1 dp_2}{2\pi}
\nonumber \\
 \int_0^\infty dt_1
\exp\left(-\frac{p_1^2+p_2^2}{2 k_f^2}+W\right).
\nonumber
\end{eqnarray}

The exponent here is represented as
\begin{eqnarray}
\exp\left[-\frac{1}{2}
(\begin{array}{cc}
p_1 & p_2 \end{array})
\hat L \left(
\begin{array}{c}
p_1 \\ p_2 \end{array} \right) \right],
\nonumber
\end{eqnarray}
where $\hat L$ is the matrix with the components
\begin{eqnarray}
L_{11}=
k_f^{-2} +2\nu (t_1+t_3)+(2/3) \nu \Sigma^2(t_1^3+t_3^3) ,
\nonumber \\
L_{12}=L_{21}=
-\nu \Sigma (t_1^2+t_3^2) ,
\nonumber \\
L_{22}=k_f^{-2} +2\nu (t_1+t_3).
\nonumber
\end{eqnarray}
Determinant of $\hat L$ is
\begin{eqnarray}
\det \hat L=
k_f^{-4} + k_f^{-2}4 \nu  (t_1+t_3)
\nonumber \\
+ k_f^{-2}(2/3) \nu  \Sigma ^2 (t_1^3+t_3^3)
+ 4 \nu ^2 (t_1+t_3)^2
\nonumber \\
 +(1/3) \nu ^2 \Sigma ^2 t_1^4
+ (4/3) \nu ^2 \Sigma ^2 t_1^3 t_3
\nonumber \\
- 2 \nu ^2 \Sigma ^2 t_1^2 t_3^2
 + (4/3) \nu ^2 \Sigma ^2 t_1 t_3^3 + (1/3) \nu ^2 \Sigma ^2 t_3^4.
\label{determ}
\end{eqnarray}
Taking the integral over $p_1,p_2$, one finds
\begin{eqnarray}
 \langle \varpi (t, \bm x) \varpi(0,\bm y)\rangle_0
={2\epsilon}k_f\frac{\partial}{\partial k_f}
 \nonumber \\
 \int_0^\infty dt_1 \exp[-\alpha(t_1+t_3)]
 (\det \hat L)^{-1/2}
\nonumber \\
\exp\left[-\frac{1}{2}(x_1-\Sigma t_3 x_2-y_1+\Sigma t_1y_2)^2 \lfloor p_1^2 \rfloor \right.
\nonumber \\
- \frac{1}{2}(x_2-y_2)^2  \lfloor p_2^2 \rfloor
-(x_1-\Sigma t_3 x_2-y_1
\nonumber \\ \left.
+\Sigma t_1y_2) (x_2-y_2)  \lfloor p_1 p_2 \rfloor
\vphantom{\frac{1}{2}} \right],
\label{pair1}
\end{eqnarray}
where
\begin{eqnarray}
\lfloor p_1^2 \rfloor= (\det \hat L)^{-1}
[k_f^{-2} +2\nu (t_1+t_3)]
\nonumber \\
\lfloor p_2^2 \rfloor= \frac{1}{\det \hat L}
\left[k_f^{-2} +2\nu (t_1+t_3)+\frac{2}{3} \nu \Sigma^2(t_1^3+t_3^3) \right],
\nonumber \\
\lfloor p_1 p_2 \rfloor= (\det \hat L)^{-1}
\nu \Sigma (t_1^2+t_3^2).
\nonumber
\end{eqnarray}
As it should be, the correlation function (\ref{pair1}) depends solely on $x_1-y_1$.

First of all, we calculate the second moment $\langle \varpi^2 \rangle$. Taking $t=0, \bm x=0, \bm y=0$, we find from Eq. (\ref{pair1})
\begin{eqnarray}
 \langle \varpi^2 \rangle
={2\epsilon}k_f\frac{\partial}{\partial k_f}
 \int_0^\infty dt_1 \exp(-2\alpha t_1)
 (\det \hat L)^{-1/2},
\label{pair7}
\end{eqnarray}
where one should substitute $t_3=t_1$. As we will see, $t_1\sim \tau_\star$. Neglecting the terms of the order of $\nu k_f^2 t_1\ll 1$ we find in the main approximation from Eq. (\ref{determ})
\begin{eqnarray}
\det \hat L\approx
k_f^{-4}  + (4/3)k_f^{-2} \nu  \Sigma ^2 t_1^3.
\nonumber
\end{eqnarray}
Substituting the expression into Eq. (\ref{pair7}), one finds after integrating over $t_1$
\begin{equation}
\langle \varpi^2 \rangle
=2\epsilon k_f^2 \tau_\star \frac{2}{\sqrt \pi}\left(\frac{4}{3}\right)^{2/3}
\Gamma\left(\frac{1}{3}\right) \Gamma\left(\frac{7}{6}\right).
\label{pair8}
\end{equation}
At the calculation one substituted $\exp(-2\alpha t_1)$ by unity since $\alpha \tau_\star \ll1$.

Consider the case, where $\bm x=\bm y=0$ and $t\gg\tau_\star$. Then
\begin{eqnarray}
 \langle \varpi (t, \bm 0) \varpi(0,\bm 0)\rangle_0
 \nonumber \\
={2\epsilon}k_f\frac{\partial}{\partial k_f}
 \int_0^\infty dt_1 \frac{\exp[-\alpha(t_1+t_3)]}
 {(\det \hat L)^{1/2}},
\label{puir1}
\end{eqnarray}
where $t_3=t+t_1$. Then $t_1\sim t$ and
\begin{equation}
\det \hat L\approx
k_f^{-2}(2/3) \nu  \Sigma ^2 [t_1^3+(t+t_1)^3].
\nonumber
\end{equation}
The approximation is valid provided $t\ll (\nu k_f^2)^{-1}$. Since $\alpha \lesssim \nu k_f^2$, we, again, can substitute $\exp(-2\alpha t_1)$ by unity in Eq. (\ref{puir1}). Then one obtains
\begin{eqnarray}
 \langle \varpi (t, \bm 0) \varpi(0,\bm 0)\rangle
 =\sqrt{6}\, c_1 \frac{\epsilon k_f^2 \tau_\star^{3/2}}{t^{1/2}},
 \label{pair9}
 \end{eqnarray}
 where $c_1$ is the following numerical factor
 \begin{equation}
 c_1=\int_{0}^{\infty}\frac{ds}{\sqrt{(1+s)^3+s^3}}\approx 1.6969  .
 \nonumber
 \end{equation}
The expression (\ref{pair9}) is correct provided $\tau_\star\ll t\ll (\nu k_f^2)^{-1}$, at $t\sim \tau_\star$ it turns to Eq. (\ref{pair8}).

If $t=0$, then the pair correlation function depends on $\bm x-\bm y$. Therefore one can take $\bm y=0$ to obtain
\begin{eqnarray}
 \langle \varpi (0, \bm x) \varpi(0,\bm 0)\rangle_0
={2\epsilon}k_f\frac{\partial}{\partial k_f}
 \int_0^\infty dt_1 \frac{\exp(- B)}{(\det \hat L)^{1/2}},
\label{pairj1}
\end{eqnarray}
where
\begin{eqnarray}
\det \hat L=
k_f^{-4} + 8 k_f^{-2} \nu t_1 + (4/3)k_f^{-2}\nu  \Sigma ^2 t_1^3
\nonumber \\
+ 8 \nu ^2 t_1^2
+ (4/3) \nu ^2 \Sigma ^2 t_1^4,
\label{determ2} \\
B=\frac{1}{2}(x_1-\Sigma t_1 x_2)^2 \lfloor p_1^2 \rfloor
+\frac{1}{2}x_2^2  \lfloor p_2^2 \rfloor
\nonumber \\
+(x_1-\Sigma t_1 x_2) x_2  \lfloor p_1 p_2 \rfloor,
\nonumber
\end{eqnarray}
that is
\begin{eqnarray}
B= (2\det \hat L)^{-1}\left\{[k_f^{-2} +4\nu  t_1] x_1^2 \right.
\nonumber \\
-2\Sigma t_1[k_f^{-2} +2\nu  t_1]x_1 x_2
+[k_f^{-2} +4\nu t_1
\nonumber \\ \left.
+\Sigma^2 t_1^2 k_f^{-2}
+(4/3)\nu \Sigma^2 t_1^3] x_2^2
\right\}.
\label{parj2}
\end{eqnarray}
We will consider different regions corresponding to different characteristic values of $t_1$. In any case, we assume $\alpha t_1\ll1$, therefore the omitted in Eq. (\ref{pairj1}) the factor $\exp(-2\alpha t_1)$.

Let us consider the case $x_1=0$. If $k_f |x_2|\gg 1$ then the expression (\ref{pairj1}) is exponentially small. That is why we analyze the opposite limit. If $k_f |x_2|\ll (\Sigma \tau_\star)^{-1}$ then the dependence of $\langle \varpi (0, \bm x) \varpi(0,\bm 0)\rangle$ on $x_2$ is negligible and we return to the second moment (\ref{pair8}). Therefore we concentrate on the region
\begin{equation}
1 \gg k_f |x_2|\gg (\Sigma \tau_\star)^{-1},
\label{region1}
\end{equation}
where a power behavior is realized. There are two different regions of integration over $t_1$ in the expression (\ref{pairj1}): small $t_1$ and large $t_1$, supplying the leading contributions to $\langle \varpi (0, \bm x) \varpi(0,\bm 0)\rangle$.

In the first region (small $t_1$) the dominant term in the curly brackets in Eq. (\ref{parj2}) is $\Sigma^2 t_1^2 k_f^{-2} x_2^2$ and we find from Eqs (\ref{determ2},\ref{parj2}) in the main approximation
\begin{equation}
\det \hat L = k_f^{-4}, \quad
B = \frac{1}{2} \Sigma^2 t_1^2 k_f^2 x_2^2.
 \label{small}
 \end{equation}
Therefore we find the characteristic $t_1$, $t_1\sim (\Sigma k_f x_2)^{-1}$. Then we obtain from Eq. (\ref{region1})
\begin{equation}
\Sigma^{-1}\ll t_1 \ll \tau_\star.
\label{parj3}
\end{equation}
The inequalities (\ref{parj3}) justify the expressions (\ref{small}). Substituting the expressions (\ref{small}) into Eq. (\ref{pairj1}), one finds the small $t_1$ contribution:
\begin{equation}
 \langle \varpi (0, \bm x) \varpi(0,\bm 0)\rangle_s
 =\sqrt{2\pi}\frac{2\epsilon k_f}{\Sigma |x_2|},
 \label{pair4s}
 \end{equation}

The second contribution stems from the region of large $t_1$, where the leading contributions to the determinant (\ref{determ2}) and to the quantity (\ref{parj2}) are
\begin{eqnarray}
\det \hat L= \frac{4}{3 k_f^2} \nu \Sigma^2 t_1^3,
\quad
B=\frac{3}{8}(\nu t_1)^{-1} x_2^2.
\label{large}
\end{eqnarray}
Therefore the characteristic $t_1$ is estimated as $t_1\sim \nu^{-1} x_2^2$ and we find from Eq. (\ref{region1})
\begin{equation}
(\nu k_f^2)^{-1}\gg t_1 \gg \tau_\star.
\label{parj7}
\end{equation}
The inequalities (\ref{parj7}) justify the expressions (\ref{large}). Substituting the expressions (\ref{large}) into Eq. (\ref{pairj1}), one finds the same expression (\ref{pair4}). Thus, at $x_1=0$ the answer (\ref{pair4}) should be doubled:
\begin{equation}
 \langle \varpi (0, \bm x) \varpi(0,\bm 0)\rangle
 =\sqrt{2\pi}\frac{4\epsilon k_f}{\Sigma |x_2|},
 \label{pair4}
 \end{equation}
 At the boundary $k_f|x_2|\sim (\Sigma \tau_\star)^{-1}$ the expression (\ref{pair4}) matches the expression for the second moment (\ref{pair8}).

Now we consider the case $x_1\neq 0$, $x_2=0$. If $k_f |x_1|\ll1$ then we return to the second moment (\ref{pair8}) for the correlation function (\ref{pairj1}). We will analyze the region
\begin{equation}
1\ll k_f |x_1|\ll \Sigma (\nu k_f^2)^{-1}
\label{region2}
\end{equation}
where the correlation function (\ref{pairj1}) possesses a power behavior. By analogy with the above analysis, one can consider contributions from small and large $t_1$. Since $k_f |x_1|\gg1$, the integrand in  (\ref{pairj1}) becomes exponentially small at small $t_1$. Therefore the contribution is negligible.

Now we pass to the contribution of large $t_1$, where the leading contributions to the determinant (\ref{determ2}) and to the quantity (\ref{parj2}) are
\begin{eqnarray}
\det \hat L= \frac{4}{3 k_f^2} \nu \Sigma^2 t_1^3,
\quad
B=\frac{3}{8}
\frac{x_1^2}{\nu \Sigma^2 t_1^3}.
\label{pairb}
\end{eqnarray}
We conclude that the characteristic value of $t_1$ is $t_1\sim x_1^{2/3}(\nu \Sigma^2)^{-1/3}$. Then we obtain from the inequalities (\ref{region2}) the same region (\ref{parj7}) for $t_1$. The inequalities (\ref{parj7}) justify the expressions (\ref{pairb}). Substituting the expressions (\ref{pairb}) into Eq. (\ref{pairj1}) and taking the resulting integral over $t_1$, one finds
\begin{equation}
\langle \varpi (0, \bm x) \varpi(0,\bm 0)\rangle
=\Gamma\left(\frac{7}{6}\right)\frac{2^{3/2} 3^{1/3} \epsilon k_f }{\Sigma^{2/3}\nu^{1/3}|x_1|^{1/3}}.
\label{pair5}
\end{equation}
At the boundary value $x_1\sim k_f^{-1}$ the expression (\ref{pair5}) matches the expression for the second moment (\ref{pair8}).

\begin{figure}
\includegraphics[width=\columnwidth]{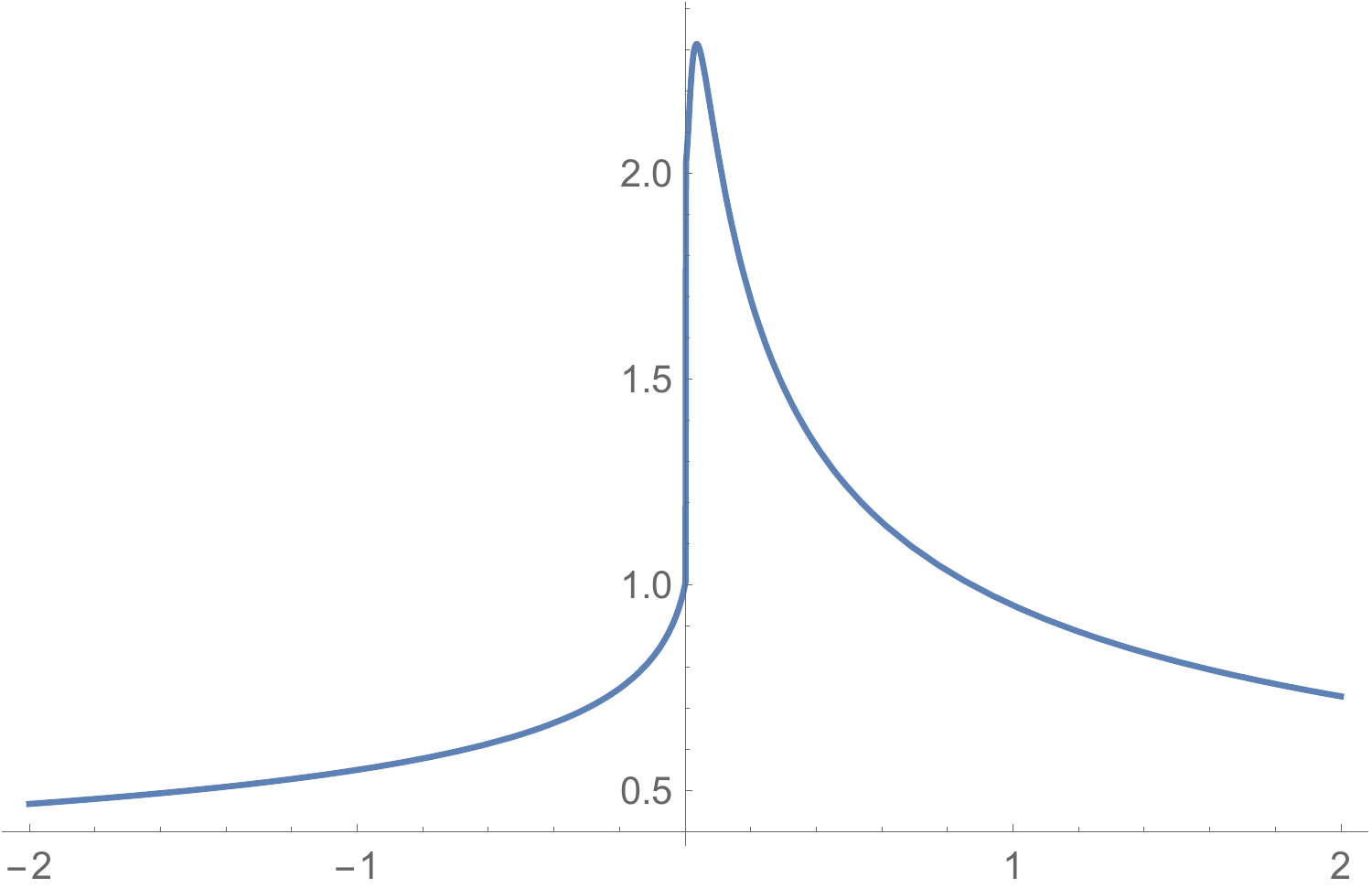}
\caption{Plot of the function $f(u)$ (\ref{pahr2}).}
\label{fig:paircorr}
\end{figure}

Now we generalize the above analysis, assuming that both, $x_1$ and $x_2$, are nonzero and are determined by the inequalities (\ref{region1},\ref{region2}). Then the contribution to $\langle \varpi (0, \bm x) \varpi(0,\bm 0)\rangle$ from small $t_1$ is absent (exponentially small) and we concentrate on the region of large $t_1$, determined by the same conditions (\ref{parj7}). Then the leading contributions to the determinant (\ref{determ2}) and to the quantity (\ref{parj2}) are
\begin{equation}
\det \hat L= \frac{4}{3 k_f^2} \nu \Sigma^2 t_1^3,
\quad
B=\frac{3}{8}
\frac{(x_1-\Sigma t_1 x_2)^2}{\nu \Sigma^2 t_1^3}.
\label{pairg}
\end{equation}
Eq. (\ref{pairg}) is a generalization of Eqs. (\ref{large},\ref{pairb}).

Substituting the expressions (\ref{pairg}) into Eq. (\ref{pairj1}) and passing to the dimensionless variable $\vartheta$, $\vartheta=\nu t_1/x_2^2$, one obtains
\begin{eqnarray}
\langle \varpi (0, \bm x) \varpi(0,\bm 0)\rangle
=\sqrt{2\pi}\frac{2\epsilon k_f}{\Sigma |x_2|}
f\left(\frac{\nu x_1}{\Sigma x_2^3}\right),
\label{pahr1} \\
f(u)=\frac{\sqrt 3}{2\sqrt{2\pi}}
\int_0^\infty \frac{d\vartheta}{\vartheta^{3/2}}
\exp\left[-\frac{3}{8\vartheta}
\left(1-\frac{u}{\vartheta}\right)^2\right].
\label{pahr2}
\end{eqnarray}
Here $f(-0)=1$ and we reproduce Eq.  (\ref{pair4s}). If $|u|\gg 1$, then
\begin{equation}
f\approx \frac{3^{1/3}\Gamma(7/6)}{\sqrt \pi |u|^{1/3}},
\nonumber
\end{equation}
reproducing Eq. (\ref{pair5}). The function $f(u)$ is plotted in Fig. \ref{fig:paircorr}. One should be careful with small $u$: the function $f(u)$ experiences the jump at $u=0$, that is observed in Fig. \ref{fig:paircorr}. The jump is explained by a finite contribution to the integral in Eq. (\ref{pahr2}) at small positive $u$ from the region $\vartheta\sim u$, at negative $u$ such contribution is absent. However, in accordance with Eqs. (\ref{region1},\ref{region2}) $|u|$ cannot be smaller than $\nu k_f^2/\Sigma$. Thus the region $|u| \lesssim \nu k_f^2/\Sigma$ is unachievable in (\ref{pahr1}),(\ref{pahr2}).

\section{Third moment of vorticity}
\label{sec:third}

The leading contribution to the third moment of vorticity is determined by the expression (\ref{generr}). It is instructive to introduce Fourier transformations for the objects figuring in the expression
\begin{eqnarray}
{ F}(t, \bm q)
=\int d^2 x \exp(-i \bm q \bm x) \langle \varpi (t, \bm 0) \varpi(0,\bm x)\rangle_0,
\label{ganer1} \\
\Phi(t, \bm k)
=\int d^2 x \exp(-i \bm k \bm x) \langle \varpi (t, \bm 0) \psi(0,\bm x)\rangle_0,
\label{ganer2} \\
{ G}(t, \bm p)
=\int d^2 x \exp(i \bm p \bm x) \langle \varpi (t, \bm 0) \mu(0,\bm x)\rangle_0.
\label{ganer3}
\end{eqnarray}
In terms of the quantities (\ref{ganer1},\ref{ganer2},\ref{ganer3}), the expression (\ref{generr}) is rewritten as
\begin{eqnarray}
\langle \varpi^3 \rangle
=6 \int_0^{\infty} dt\, \int \frac{d^2 q\, d^2k}{(2\pi)^4}
 (q_1 k_2-q_2 k_1)
 \nonumber \\
{ F}(t, \bm q) { \Phi}(t,\bm k)
{ G}(t,\bm q+\bm k).
\label{generq}
\end{eqnarray}
Note the wave vector conservation law.

One can express the quantities (\ref{ganer1},\ref{ganer2},\ref{ganer3}) via the Green function and the pair correlation function. We find
\begin{eqnarray}
{ G}(t,\bm p)
=\int \frac{d^2 k}{(2\pi)^2}
{\mathcal G}(t,\bm k,\bm p)
\nonumber \\
=\theta(t)
\exp\left(-\alpha t-\nu \bm p^2 t
+\nu\Sigma p_2 p_1 t^2-\frac{1}{3}\nu \Sigma^2 p_1^2 t^3
\right).
\label{zerot1}
\end{eqnarray}
Note, that one can rewrite Eq. (\ref{gref2}) as
\begin{eqnarray}
{\mathcal G}(t,\bm k,\bm q)
=(2\pi)^2 \delta(k_1-q_1)
\nonumber \\
\delta\left(k_2-q_2+\Sigma k_1t\right)
{ G}(t,\bm q).
\label{zerot2}
\end{eqnarray}
We find for the object (\ref{ganer2})
\begin{eqnarray}
{ F}(t,\bm q)
=\int \frac{d^2 k}{(2\pi)^2}
{\mathcal F}(t,\bm k,\bm q)
\nonumber \\
= 2\pi \frac{2\epsilon}{k_f^2} \int_0^\infty dt_1 (q_1^2+q_3^2)
\exp\left(-\frac{q_1^2+q_3^2}{2 k_f^2}\right)
\nonumber \\
{ G}(t+t_1,q_1,q_3) { G}(t_1,-q_1,-q_3)
\nonumber
\end{eqnarray}
where $q_3=q_2+\Sigma q_1 t_1$. Assuming $t>0$, we find
\begin{eqnarray}
{ F}(t,\bm q)
= 2\pi \frac{2\epsilon}{k_f^2} \int_0^\infty dt_1 (q_1^2+q_3^2)
\exp\left(-\frac{q_1^2+q_3^2}{2 k_f^2}\right)
\nonumber \\
\exp\left[-(\alpha +\nu q_1^2  +\nu q_3^2) (t_1+t_3)
\vphantom{\frac{1}{2}} \right.
\nonumber \\ \left.
+\nu\Sigma  q_1 q_3 (t_1^2+t_3^2)
-\frac{1}{3}\nu \Sigma^2 q_1^2 (t_1^3+t_3^3)
\right],
\label{fouri5}
\end{eqnarray}
where $t_3=t+t_1$, $q_3=q_2+\Sigma q_1 t_1$. Analogously for $t>0$ we find
\begin{eqnarray}
\Phi(t,\bm k)
=2\pi \frac{2\epsilon}{k_f^2} \int_0^\infty dt_2
\frac{k_1^2+k_3^2}{k_1^2+k_2^2}
\exp\left(-\frac{k_1^2+k_3^2}{2 k_f^2}\right)
\nonumber \\
\exp\left[-(\alpha +\nu k_1^2  +\nu k_3^2) (t_2+t_4)
\vphantom{\frac{1}{2}} \right.
\nonumber \\ \left.
+\nu\Sigma k_3 k_1 (t_2^2+t_4^2)
-\frac{1}{3}\nu \Sigma^2 k_1^2 (t_2^3+t_4^3)
\right],
\label{fouri6}
\end{eqnarray}
where $t_4=t+t_2$, $k_3=k_2+\Sigma k_1 t_2$.

\begin{figure}
\includegraphics[width=\columnwidth]{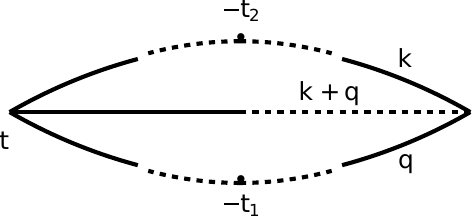}
\caption{Feynman diagram representing first contribution to the third moment.}
\label{fig:corpum}
\end{figure}

To find the leading contribution to the third moment, one has to substitute the expressions (\ref{zerot1},\ref{fouri5},\ref{fouri6}) into Eq. (\ref{generq}) and to calculate the resulting integral. Note that the contribution can be represented by Feynman diagram depicted in Fig. \ref{fig:corpum}. The designations here are the same as in Fig. \ref{fig:corpum}: the combined solid-dashed line represents the Green function (\ref{fouri1}), the bullet represents the pumping. Further we pass to integration over the components $k_1, k_3, q_1, q_3$ in Eq. (\ref{generq}). Then one should substitute
\begin{equation}
q_2=q_3-\Sigma q_1 t_1, \quad
k_2=k_3-\Sigma k_1 t_2,
\label{logsi}
\end{equation}
into the expressions (\ref{fouri5},\ref{fouri6}) for $F,\Phi$. Note that the components $k_1, k_3, q_1, q_3$ are restricted by pumping, and therefore they cannot be much larger than $k_f$.

Further we examine the main contribution to the integral (\ref{generq}). As we will see, the contribution originates from the region of integration where $t, t_1 \gg \tau_*, t_2$, $k_1,q_1\ll k_f$ whereas $k_3,q_3\sim k_f$. Therefore one can neglect $k_1,q_1$ in comparison with $k_3,q_3$. Next, we neglect $t_2$ in comparison with $t,t_1$ and keep solely the third order terms in $t,t_1$ in the exponents in Eqs. (\ref{zerot1},\ref{fouri5},\ref{fouri6}) in terms of the components $(k_1,k_3,q_1,q_3)$, see Eq. (\ref{logsi}). We neglect the terms with $\alpha$ as well. Then
\begin{eqnarray}
{ G}(t,\bm k+\bm q)
\to\theta(t)\exp\left[ -\nu \Sigma^2 t_1^2 t q_1^2
\vphantom{\frac{1}{2}} \right.
\nonumber \\ \left.
-\nu\Sigma^2 t_1 t^2 q_1(q_1+k_1)
-\frac{1}{3}\nu \Sigma^2 (q_1+k_1)^2 t^3 \right],
\label{logsi1}
\end{eqnarray}
\begin{eqnarray}
{ F}(t,\bm q)
\to 2\pi \frac{2\epsilon}{k_f^2} \int_0^\infty dt_1 q_3^2
\exp\left(-\frac{q_3^2}{2 k_f^2}\right)
\nonumber \\
\exp\left[-\frac{1}{3}\nu \Sigma^2 q_1^2 (t_1^3+t_3^3)\right],
\label{logsi2}
\end{eqnarray}
\begin{eqnarray}
\Phi(t,\bm k)
=2\pi \frac{2\epsilon}{k_f^2} \int_0^\infty dt_2
\frac{k_3^2}{k_1^2+k_2^2}
\nonumber \\
\exp\left(-\frac{k_3^2}{2 k_f^2}\right)
\exp\left[-\frac{1}{3}\nu \Sigma^2 k_1^2 t^3 \right].
\label{logsi3}
\end{eqnarray}
In the situation one can integrate over $t_2$ in Eq. (\ref{logsi3}) to obtain
\begin{equation}
\int_0^\infty dt_2
\frac{1}{k_1^2+(\Sigma t_2 k_1 -k_3)^2}
\to \pi \Sigma^{-1} k_1^{-2} \theta(k_1 k_3).
\label{logac1}
\end{equation}
The Heaviside function in Eq. (\ref{logac1}) means that the integral over $t_2$ for identical signs of $k_1,k_3$ is much larger than for different signs of $k_1,k_3$. Therefore we find from Eq. (\ref{logsi3})
\begin{eqnarray}
\Phi(t,\bm k)
\to \frac{(2\pi)^2\epsilon}{\Sigma k_f^2}
\theta(k_1k_3) \frac{k_3^2}{k_1^2}
\exp\left(-\frac{k_3^2}{2 k_f^2}\right)
\nonumber \\
\exp\left(-\frac{1}{3}\nu \Sigma^2 k_1^2 t^3\right).
\label{phiex2}
\end{eqnarray}

Substituting the expressions (\ref{logsi1},\ref{logsi2},\ref{phiex2}) into Eq. (\ref{generq})and taking the integrals over $k_3,q_3$, we find
\begin{eqnarray}
\langle \varpi^3 \rangle
=6 \epsilon^2 k_f^{2}\iint_0^{\infty} dt\, dt_1\, t_1
\int dk_1\, dq_1\, \frac{q_1}{k_1}
\nonumber \\
\exp\left\{ -\nu \Sigma^2 \left[\frac{2}{3}(t_1+t)^2 q_1^2 \right. \right.
\nonumber \\ \left. \left.
+\left(t_1 t^2+\frac{2}{3}t^3\right)q_1 k_1 +\frac{2}{3}t^3k_1^2 \right]\right\}.
\label{logac3}
\end{eqnarray}
Taking then the integral over $q_1$, one obtains
\begin{eqnarray}
\langle \varpi^3 \rangle
=-\frac{6 \epsilon^2 k_f^{2}}{\Sigma \nu^{1/2}}\iint_0^{\infty} dt\, dt_1\, t_1
\int dk_1\, \sqrt{2\pi}\frac{b}{a^{3/2}}
\nonumber \\
\exp\left\{ \nu \Sigma^2 \left[\frac{b^2}{2a}-
\frac{2}{3}t^3 \right]k_1^2\right\},
\label{logac4}
\end{eqnarray}
where $a=(4/3)(t_1+t)^3$, $b=t_1t^2 +(2/3)t^3$. Taking then the integral over $k_1$, one finds
\begin{eqnarray}
\langle \varpi^3 \rangle
=-\frac{12 \pi \epsilon^2 k_f^{2}}{\Sigma^2 \nu}\iint_0^{\infty} dt\, dt_1\, t_1
\frac{(b/a)}{\sqrt{(4/3)t^3a -b^2}}.
\label{logac5}
\end{eqnarray}

Passing to the variable $\eta=t_1/t$, one obtains the logarithmic integral
\begin{eqnarray}
\langle \varpi^3 \rangle
=-\frac{9 \pi \epsilon^2 k_f^{2}}{\Sigma^2 \nu}c_2
\int\frac{dt}{t},
\label{logac6}
\end{eqnarray}
where $c_2$ is the numerical factor
\begin{eqnarray}
c_2=\int_0^\infty d\eta \frac{\eta (3\eta+2)}{(1+\eta)^3}
(12 +36 \eta +39 \eta^2 +16 \eta^3)^{-1/2},
\nonumber \\
c_2 \approx 0.187.
\nonumber
\end{eqnarray}
Since $\eta\sim 1$, then $t_1\sim t$. Note also that
\begin{equation}
k_1, q_1\sim (\nu \Sigma^2 t^3)^{-1/2},
\label{logac7}
\end{equation}
as follows from the corresponding integrals.

The integration over $t$ in the integral (\ref{logac6}) is restricted from above by the viscous time $(\nu k_f^2)^{-1}$ since at larger times the neglected in the exponents in Eqs. (\ref{zerot1},\ref{fouri5},\ref{fouri6}) terms of the type $\nu k_3^2t$ start to work. We neglected $k_1,q_1$ in comparison with $k_f$, therefore $t\gg \tau_\star$, as it follows from Eq. (\ref{logac7}). Thus, we conclude that the time $t$ lies in the interval
\begin{equation}
(\nu k_f^2)^{-1}\gtrsim t \gtrsim \tau_\star.
\label{logac8}
\end{equation}
In terms of $q_1,k_1$ the interval (\ref{logac8}) is written as
\begin{equation}
k_f\gtrsim q_1,k_1 \gtrsim \frac{\nu k_f^2}{\Sigma}k_f,
\label{logac9}
\end{equation}
as it follows from Eq. (\ref{logac7}). Note that the integral (\ref{logac1}) is gained near
\begin{equation}
t_2\approx k_3(\Sigma k_1)^{-1}\sim k_f (\Sigma k_1)^{-1}.
\label{logac11}
\end{equation}
As it follows from Eqs. (\ref{logac7},\ref{logac11}), the ratio
\begin{equation}
t_2/t \sim (\nu k_f^2 t)^{1/2}
\nonumber
\end{equation}
is small in the interval (\ref{logac8}). As it is demonstrated in the work \cite{KL20}, the coherent vortex is formed if the condition $\alpha\lesssim \nu k_f^2$ is satisfied. Since the terms of the type $\nu k_f^2 t$ can be neglected, the terms with $\alpha$ can be neglected as well. Thus, our approximations are justified in the interval (\ref{logac8}).

The logarithmic integral (\ref{logac6}) is gained in the wide interval of $t$ determined by the conditions (\ref{logac8}). To find the integral it is enough to substitute the lower and the higher values of $t$ from the conditions (\ref{logac8}) as the limits in the logarithmic integral (\ref{logac6}) and we find
\begin{eqnarray}
\langle \varpi^3 \rangle
=-\frac{6 \pi \epsilon^2 k_f^{2}}{\Sigma^2 \nu}c_2
\ln \frac{\Sigma}{\nu k_f^2},
\label{logac10}
\end{eqnarray}
with the logarithmic accuracy. Remind that we assumed $\Sigma>0$ in our calculations. Therefore the sign minus in Eq. (\ref{logac10}) means, that the third moment has the sign, opposite to the sign of the shear rate $\Sigma$.

\section{Conclusion}
\label{sec:conclu}

We analytically investigated the pair correlation function of the fluctuating vorticity and its third moment inside the coherent vortex appearing as a consequence of the inverse cascade in two-dimensional turbulence. The particular choice of the pumping (external forcing) enabled us to obtain explicit analytic expressions for the objects. The expressions (\ref{pair8},\ref{pair9},\ref{pair4},\ref{pair5},\ref{pahr1})  are in accordance with the general analysis given in the work \cite{KLT23}. The expression (\ref{logac10}) for the third moment is in accordance with the general predictions given in the work \cite{KL23}.

One new ingredient is the logarithm in the expression (\ref{logac10}) derived with the logarithmic accuracy. It is related to the fact that the third moment is gained from the wide region of times $\tau_\star < t <(\nu k_f^2)^{-1}$, whereas the second moment is gained from the region $t\sim \tau_\star$. The existence of the wide region of times is, obviously, universal, that is independent of the concrete shape of pumping. We are going to investigate the role of the fact for higher moments of $\varpi$.

It is worth noting that there are also fluctuations with the scale of the order of the system size $L$. It was shown in \cite{Frishman1} that such fluctuations are important in analysis of velocity statistics. However for $k_f L\gg 1$ their contribution to vorticity moments is negligible.

I.V. Kolokolov's work was carried out within the framework of the scientific program of the National Center for Physics and Mathematics (the project "Physics of High Energy Densities. Stage 2023-2025"),
M.M.Tumakova thanks the Basis Foundation for its support, V.V.Lebedev thanks the support of the grant of the Ministry of Higher Education and Science of the Russian Federation, project No. 075-15-2022-1099.

\end{document}